\documentclass{ws-procs9x6}
\usepackage{epsfig}
\usepackage{wrapfig}
\def\as{\alpha_s}
\def\mz{M_Z}
\def\asz{\as(\mz)}
\def\asmz#1#2#3#4#5#6{\asz = #1\pm #2\ {\rm (stat.)}\ ^{+#4}_{-#3}\ {\rm (exp.)}\ ^{+#6}_{-#5}\ {\rm (th.)}}
\def\etjet{E_T^{\rm jet}}
\def\etajet{\eta^{\rm jet}}

\begin{document}

\title{Jet Production at HERA and Measurements of the Strong Coupling Constant $\as$}

\author{Dorian Kcira}

\address{University of Wisconsin\\
ZEUS / DESY, Notkestrasse 85,
22607 Hamburg, Germany\\
Email: dorian.kcira@desy.de}

\maketitle

\abstracts{%
Measurements of HERA that explore the parton dynamics at low Bjorken $x$ are presented
together with precise determinations of the strong coupling constant $\as$.
Calculations at next to leading order using the DGLAP evolution fail to describe the data
at low $x$ and forward jet pseudorapidities. The $\asz$ measurements at HERA
are in agreement with the world average and have very competitive errors.
}

\section{Introduction}

At HERA, forward jet production and jet-jet correlations are
expected to be sensitive to the parton dynamics at low Bjorken $x$ ($x_{\rm Bj}$).
In the first section, these type of measurements and comparison to DGLAP, BFKL
and other models are presented. For the measurements presented in
the second section, jets were selected in kinematic regions where the proton parton distribution functions (PDFs) are well
constrained and the DGLAP equations are valid.
These measurements allow precise tests of perturbative QCD (pQCD) and
the determination of the strong coupling constant, $\as$.

\section{QCD dynamics at low $x$}\label{sec:lowx}

Inclusive forward jet production was measured by the H1 Collaboration~\cite{h1_forward_jets} for events in the
kinematic region $5<Q^2<85$~GeV$^2$ and $10^{-4}<x_{\rm Bj}<4\cdot 10^{-3}$, where $Q^2$ is the virtuality of the
exchanged photon. Jets were found in the laboratory frame with
$\etjet>3.5$~GeV, $7^{\circ}<\theta_{\rm jet}<20^{\circ}$ (corresponding to
the $1.74<\etajet<2.8$~\footnote{$\eta=-\log[\tan(\theta/2)]$ is the pseudorapidity,
where $\theta$ is the polar angle.}), $x_{\rm jet}>0.035$, and $0.5<E_{\rm T,jet}^2 / Q^2<5$.
$\etjet$ is the transverse energy of the jet and $x_{\rm jet}$ is the fractional energy of
the proton taken by the jet.

\begin{figure}[ht]
\centerline{
\epsfxsize=1.5in\epsfbox{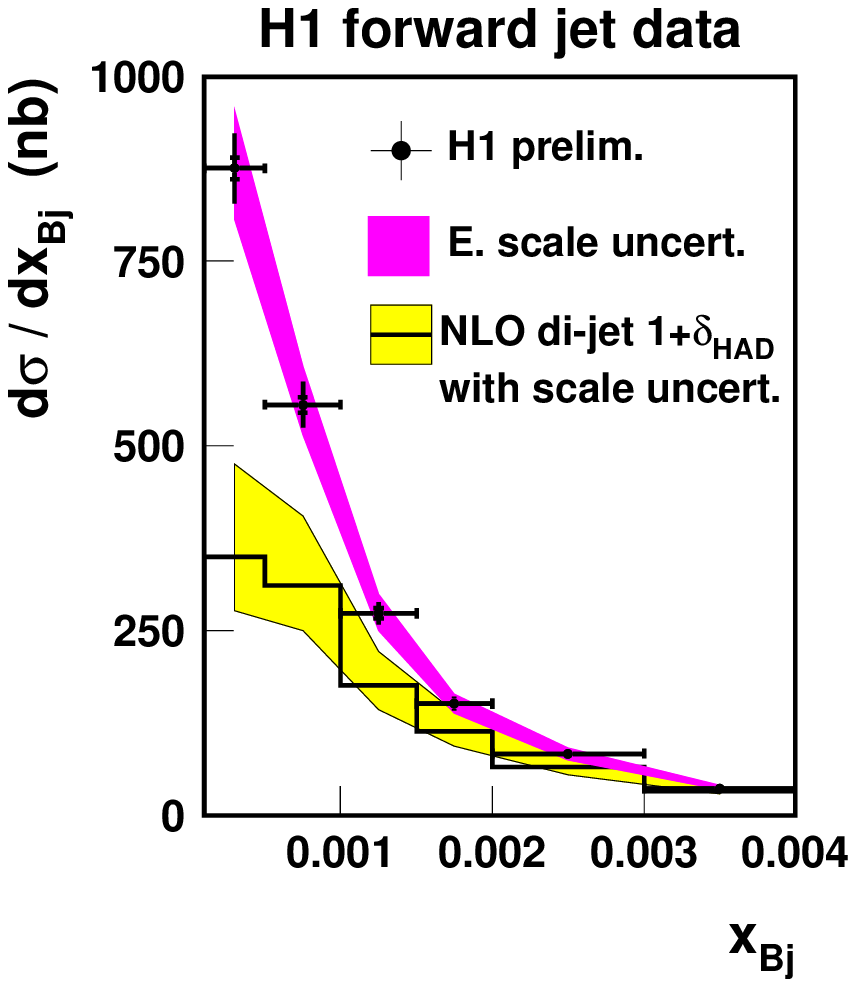}
\epsfxsize=1.5in\epsfbox{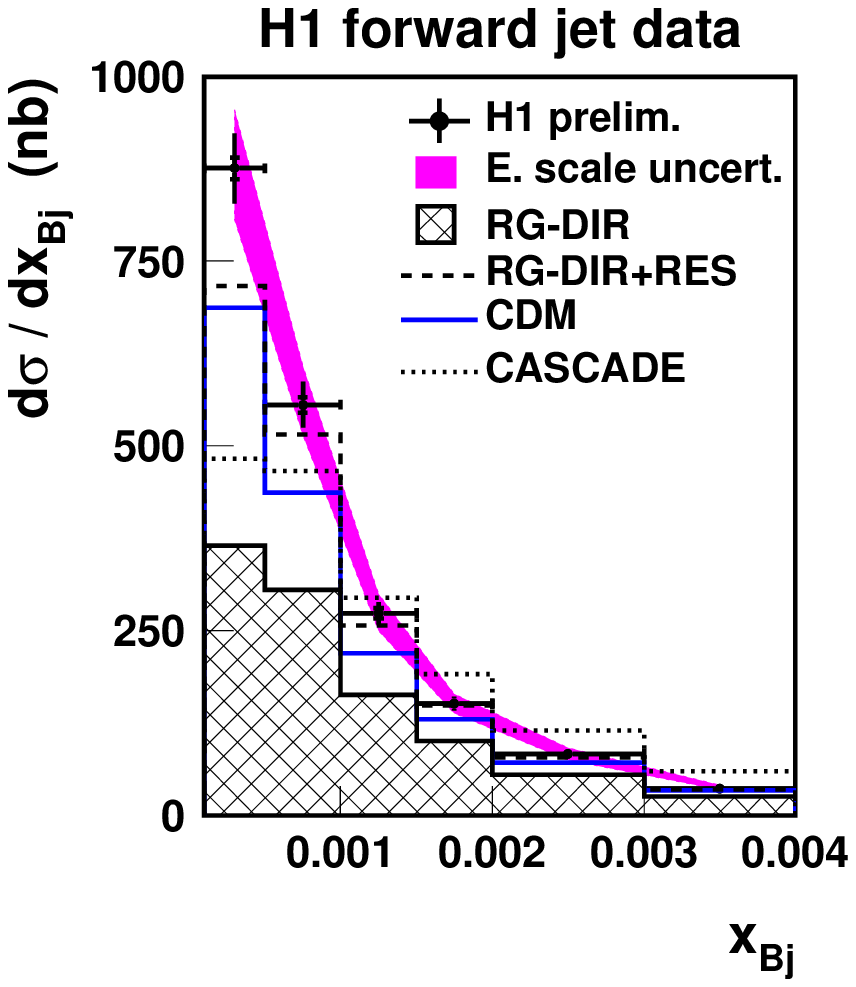}
\epsfxsize=1.65in\epsfbox{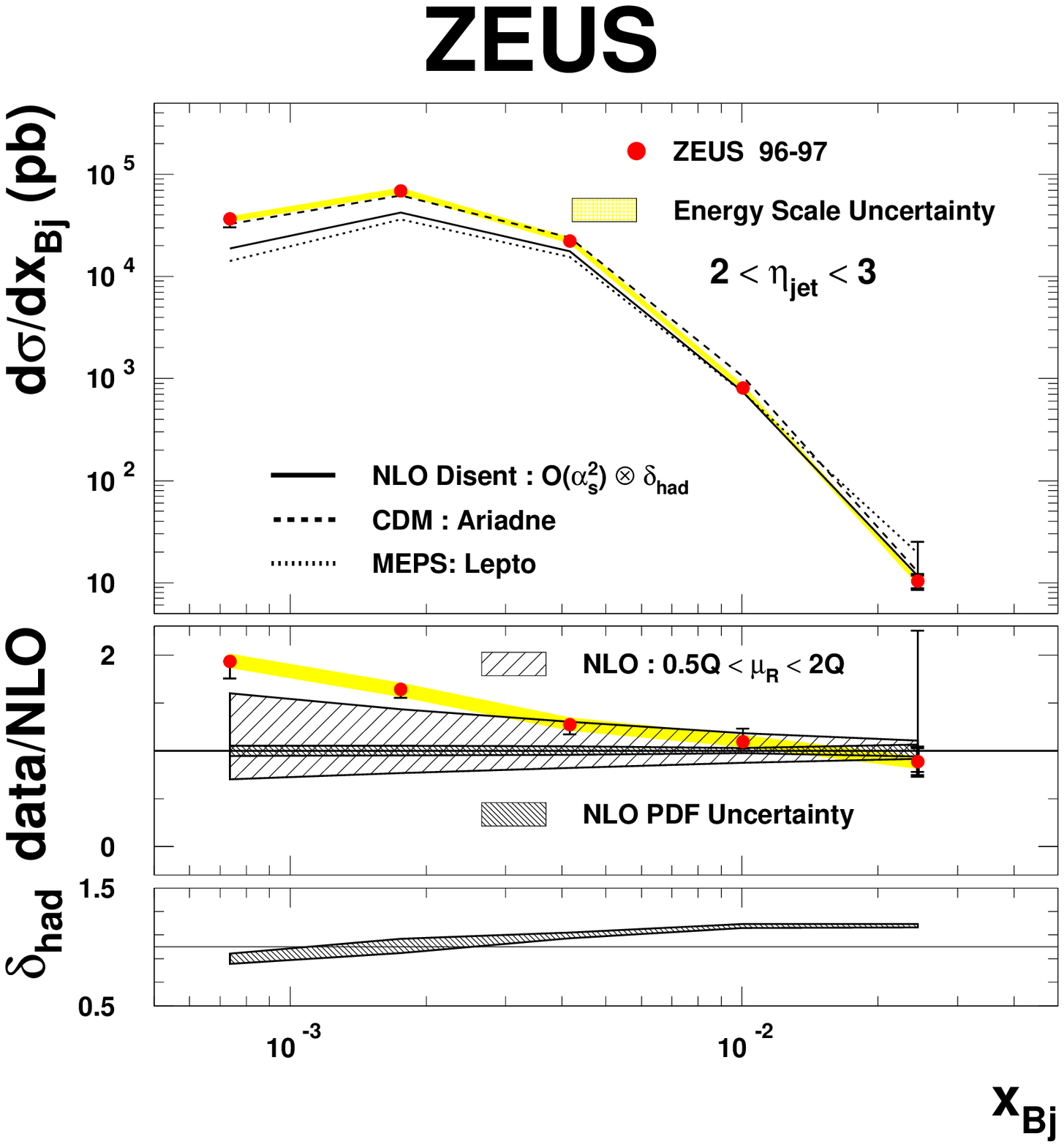}
}
\caption{%
Cross section for inclusive forward jet production as a function of $x_{\rm Bj}$ compared to
NLO calculation (left) and QCD MC models (middle).
Measured differential cross section (right) for inclusive jet production in the
forward jet rapidities as a function of $x_{\rm Bj}$.
\label{ZEUSH1fwjets}
}

\end{figure}

The measured cross section is shown in Fig.~\ref{ZEUSH1fwjets} as a function of $x_{\rm Bj}$ and
compared to the prediction of NLO calculations from DISENT
(left) and various QCD models (middle).
The DISENT calculations were performed using the CTEQ6M
parametrization of the proton PDFs.
The renormalization scale was chosen to be $\mu_R=\etjet$.
The DGLAP model with direct photon interactions alone (RG-DIR, RAPGAP)
and the NLO calculation fall below the data, especially at low $x_{\rm Bj}$.
The description of the data by RAPGAP is significantly improved
if contributions from resolved photon interactions are included (RG-DIR+RES).
The Color Dipole Model
(CDM) shows a similar behaviour to RG-DIR+RES. In addition, the CCFM
based CASCADE
MC predicts a different shape of the distribution that results in a poor description of the data.

Jet production in NC DIS has been measured by the ZEUS Collaboration~\cite{zeus_forward_jets} for $Q^2>25$~GeV$^2$, $y>0.004$,
$E^{\prime}_e>10$~GeV (where $E^{\prime}_e$ is the energy of the scattered positron), and
$\cos(\gamma_{h})<0$~\footnote{$\gamma_h$ is the hadronic angle. It corresponds,
in the Quark Parton Model, to the angle of the scattered quark.}.
Jets were selected in the laboratory frame with
$\etjet>6$~GeV, $2<\etajet<3$, and $0.5<(\etjet)^2 / Q^2<2$.

The measurement is presented in Fig.~\ref{ZEUSH1fwjets} (right) as a function of $x_{\rm Bj}$ and compared
to NLO calculations (DISENT, $\mu_R=\mu_F=Q$, CTEQ6) and to the
CDM (ARIADNE)
and MEPS (Matrix Elements + Parton Showers, LEPTO) models.
The CDM prediction gives a reasonable description of the data.
The NLO and MEPS predictions fail to describe the data in the low $x_{\rm Bj}$ region.
The uncertainty induced by the variation of the renormalization scale is large, indicating that
missing higher order or $\ln(1/x)$ terms in the calculation could be important in this region.

The H1 Collaboration has measured inclusive dijet production in DIS~\cite{h1_inclusive_dijets} in the kinematic range
$5<Q^2<100$~GeV$^2$, $10^{-4}<x<10^{-2}$, and $0.1<y<0.7$. Dijets were reconstructed in the
hadronic center-of-mass system (HCM) and selected with the requirements: $-1<\eta_{\rm jet,lab}<2.5$
and $(E_T^{\rm jet1,2})^*>7,5$~GeV. The azimuthal asymmetry is defined by:
$
S(\alpha) =%
\frac{\int_0^{\alpha} N_{\rm dijet}(\Delta \phi^*,x,Q^2){\rm d}\Delta \phi^*}%
{\int_0^{180^\circ} N_{\rm dijet}(\Delta \phi^*,x,Q^2){\rm d}\Delta \phi^*}\;,%
$
where $\Delta \phi^*$ is the azimuthal separation in the HCM frame between the two hardest transverse energy jets.

\begin{figure}[ht]
\centerline{
\epsfxsize=1.5in\epsfbox{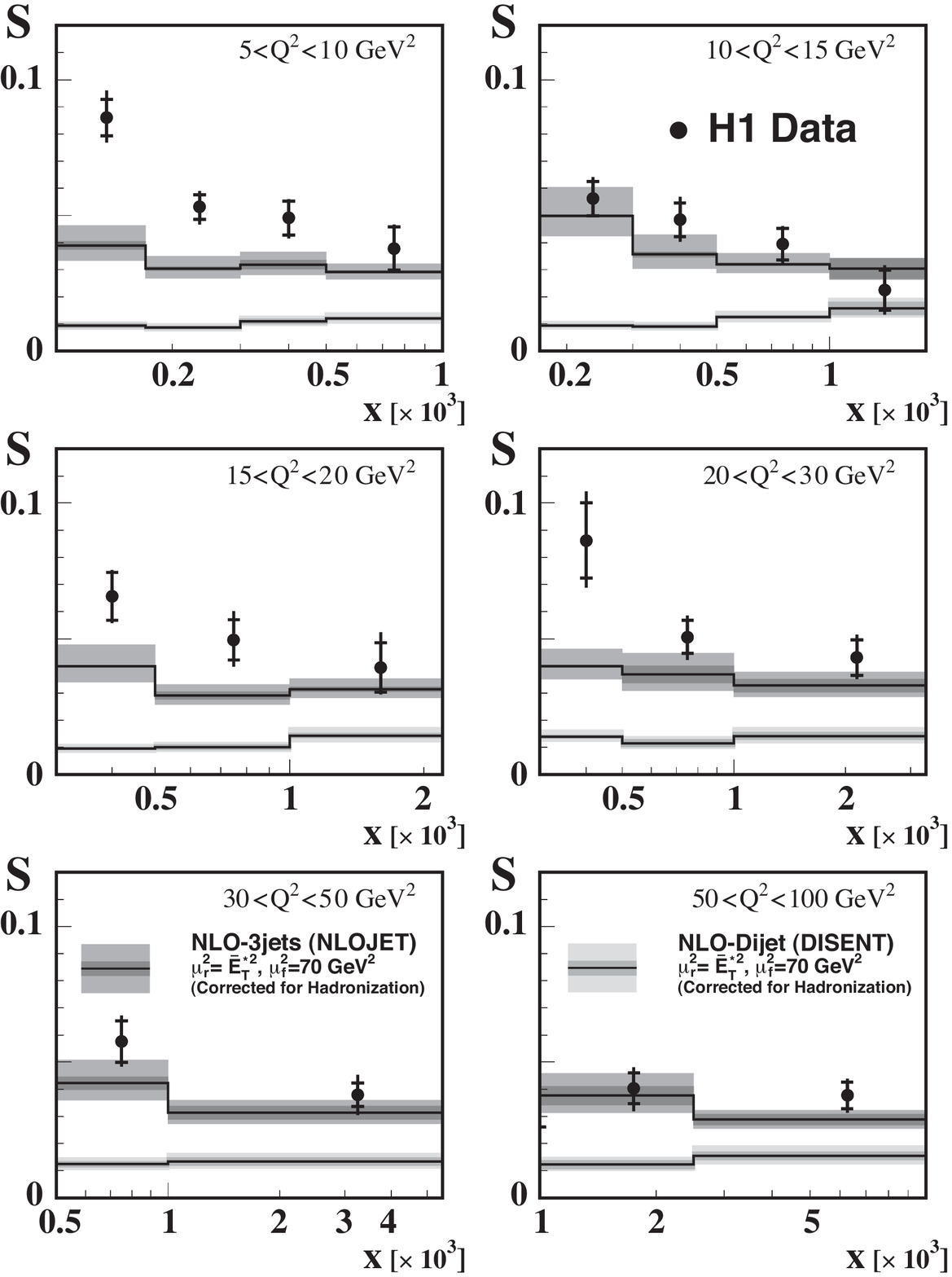}
\epsfxsize=1.5in\epsfbox{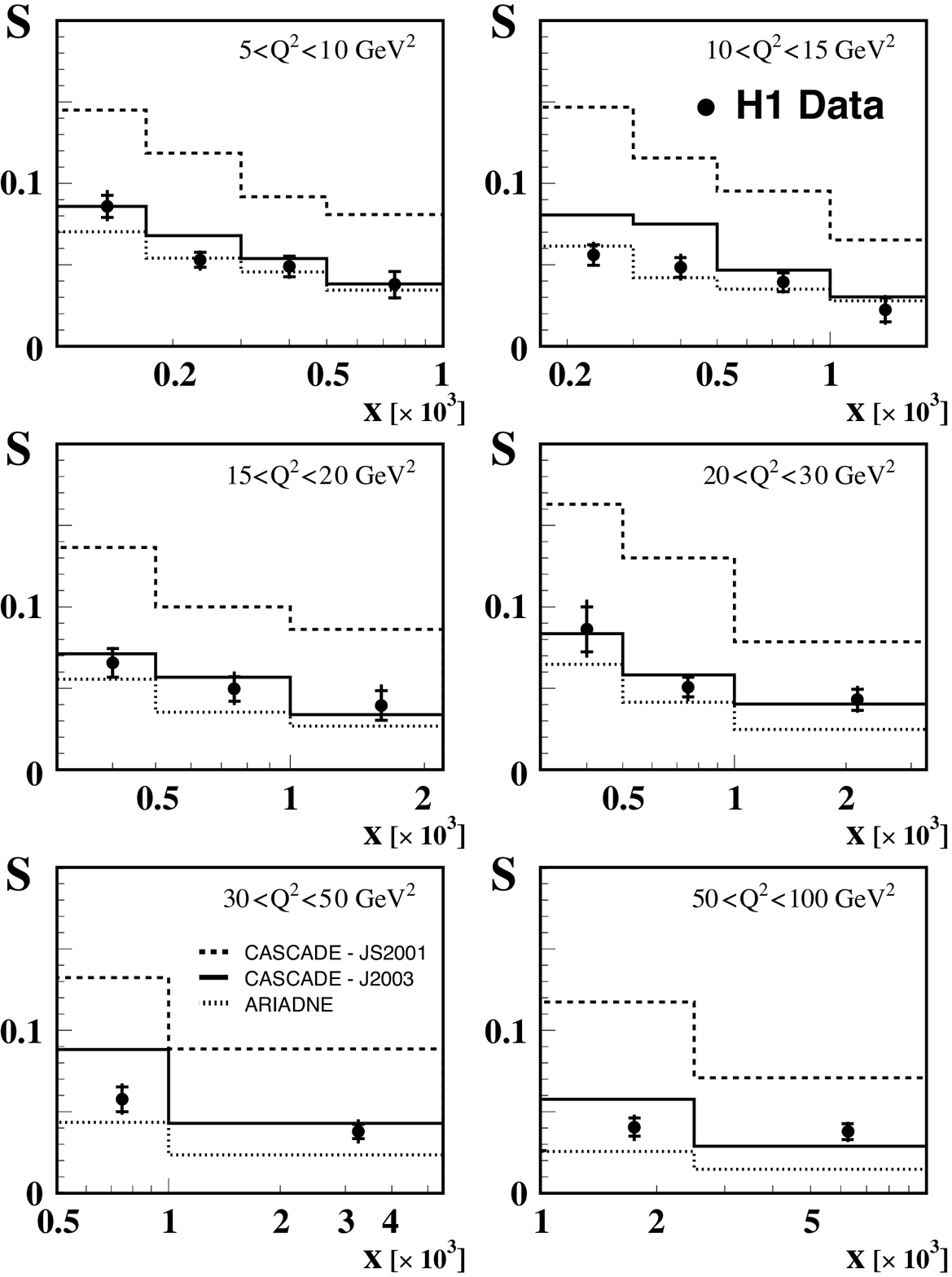}
}
\caption{
Azimuthal asymmetry $S$ for $\Delta \phi^*<120^{\circ}$ as a function of $x_{\rm Bj}$ and $Q^2$. The
data are compared to NLO predictions (left) and to the CCFM and CDM models (right).
\label{d03-160}}
\end{figure}

Figure~\ref{d03-160} presents the $S$ distribution for $\alpha=120^{\circ}$ as a function
of $x$ for different $Q^2$ compared to predictions of DGLAP NLO calculations (left) and
different models (right). The measured values of $S$ for the chosen $\alpha$  are of the order of
$5\%$ and increase with decreasing $x$. The rise is most prominent in the lowest $Q^2$ bin. The DISENT
($\mathcal{O} (\as^2)$) calculations predict no rise of $S$ with $x$ and fall below the measurements.
The NLOJET
($\mathcal{O} (\as^3)$) give a good description of the data at large $Q^2$ and large $x$
but fail to describe the strong rise towards low $x$, particularly in the lowest $Q^2$ range. A good description
of the $S$ distribution at low $x$ and low $Q^2$ is predicted from the CDM model. The CCFM based CASCADE
predictions depend strongly on the choice of the unintegrated gluon density.

\section{Precise tests of QCD and the measurement of $\as$}\label{sec:pqcd}

\begin{figure}[ht]
\centerline{
\epsfxsize=1.5in\epsfbox{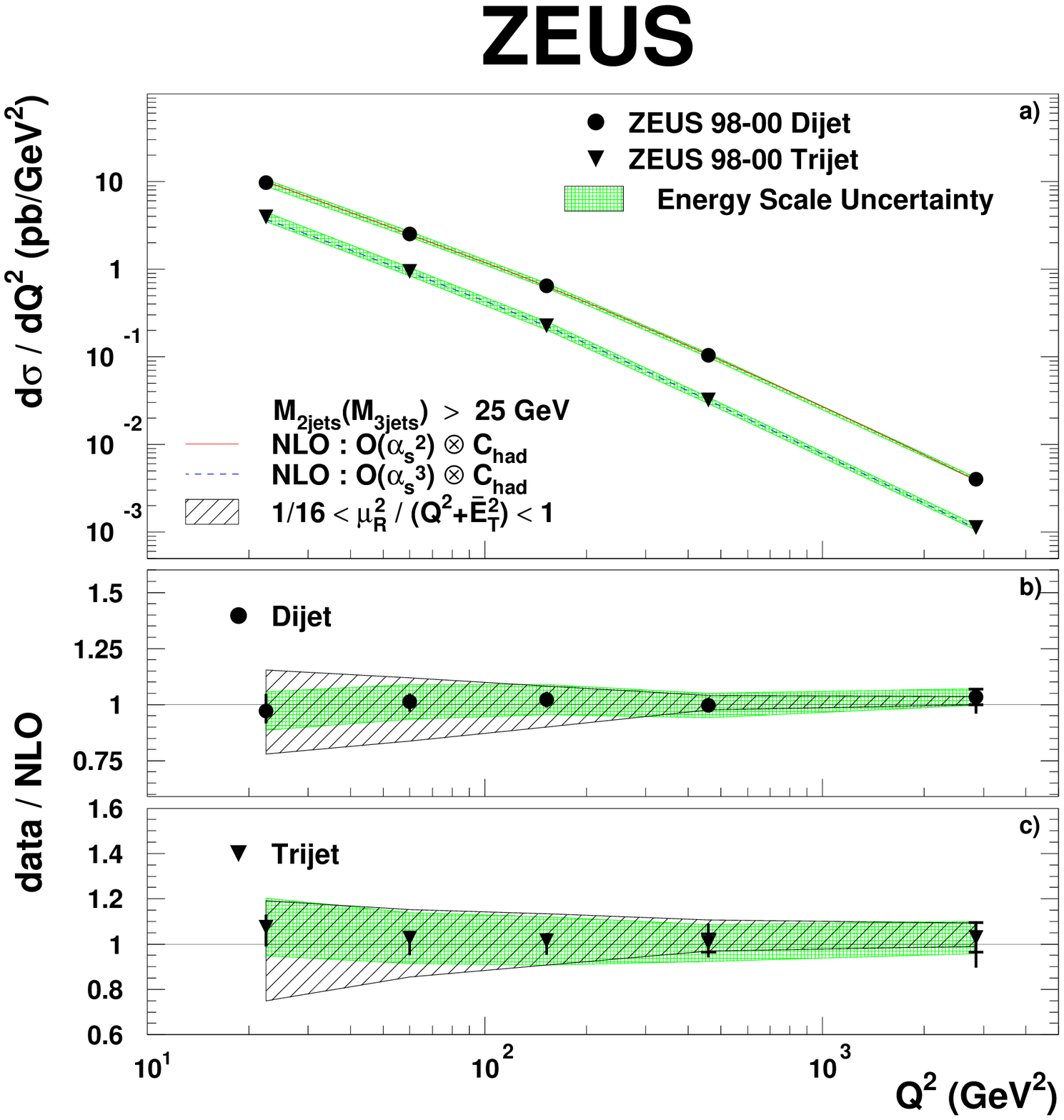}
\epsfxsize=1.5in\epsfbox{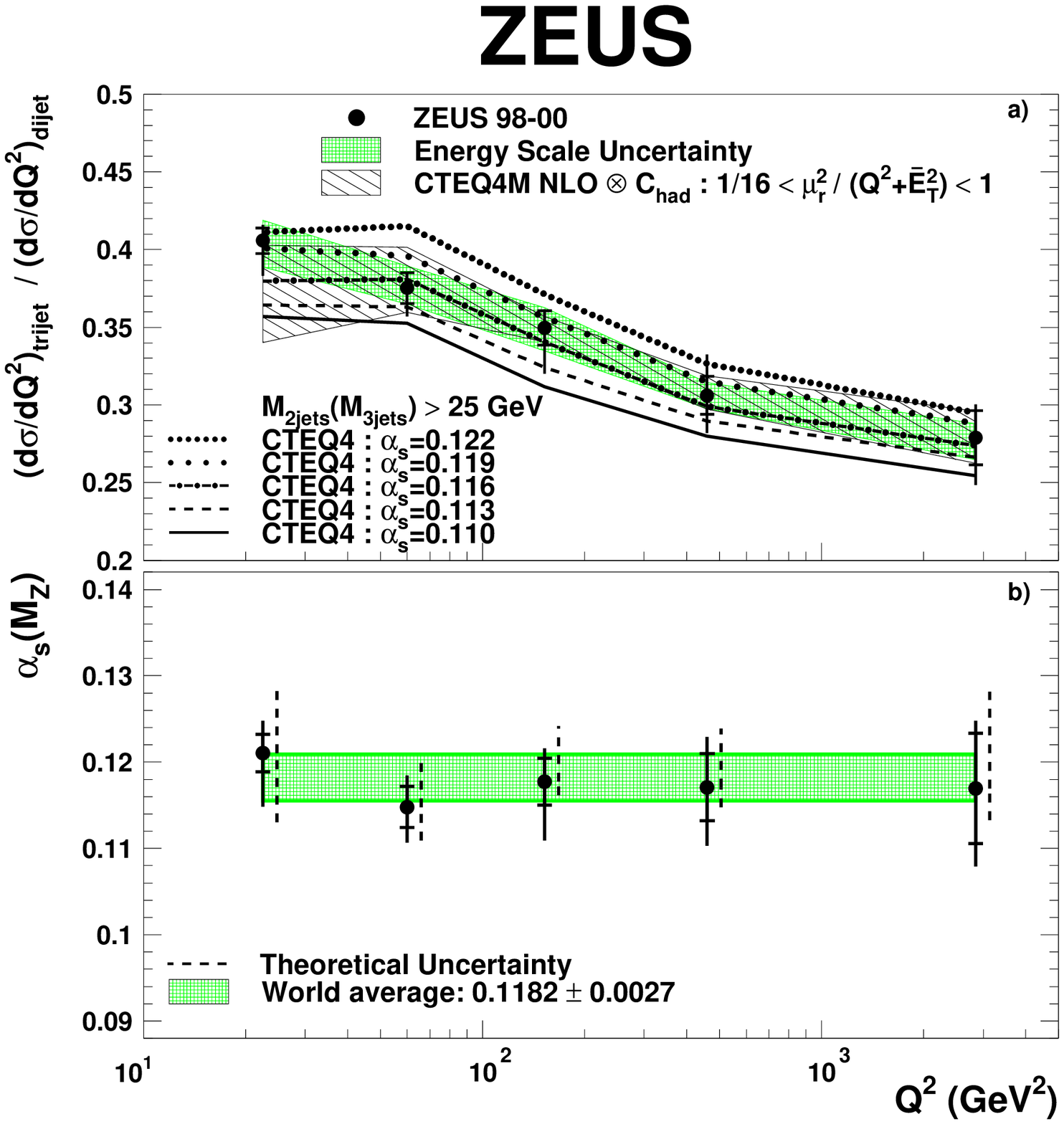}
\epsfxsize=0.95in\epsfbox{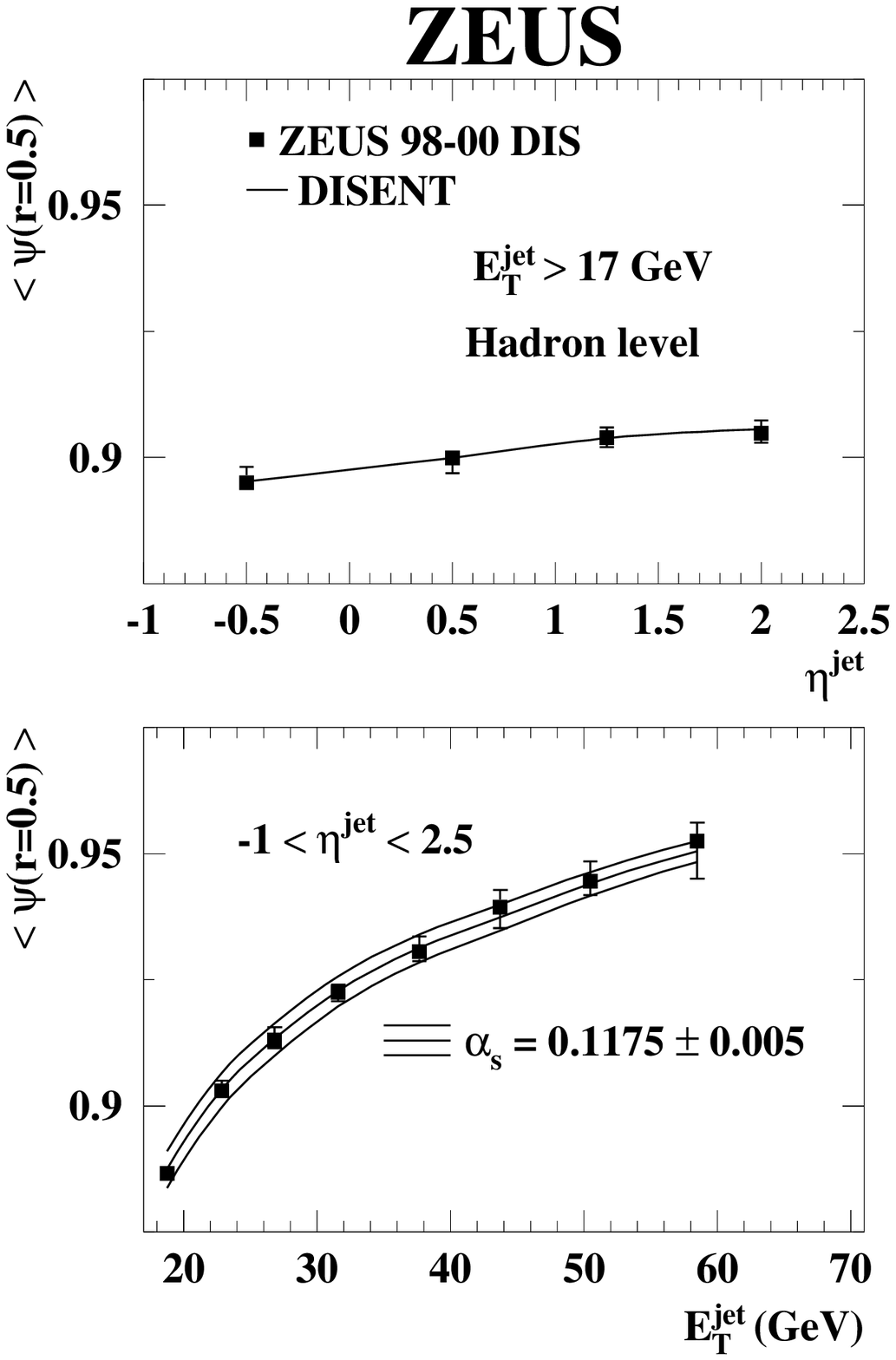}
}
\caption{
Inclusive dijet and trijet cross sections (left) and their ratio as a function of $Q^2$ (middle)
together with the extracted values of $\as$ in each $Q^2$ bin.
Mean integrated jet shape $\langle\psi(r=0.5)\rangle$ versus $\etjet$ and $\etajet$ (right)
in comparison with the expectation for three different $\as$ values.
\label{zeus_multijets_subjets}}
\end{figure}

Differential dijet and trijet cross sections have been measured by the ZEUS Collaboration~\cite{zeus_multijets}
in the kinematic range $10<Q^2<5000$~GeV$^2$ and $0.04<y<0.6$.
Jets were found in the Breit frame and dijet (trijet) events were selected with:
$-1<\eta_{\rm jet,lab}<2.5$, $E_{T\rm ,jet, Breit}>5$~GeV, and invariant mass $M_{\rm 2/3jets}>25$~GeV.

\begin{wrapfigure}[15]{r}{5.4cm}
\epsfig{file=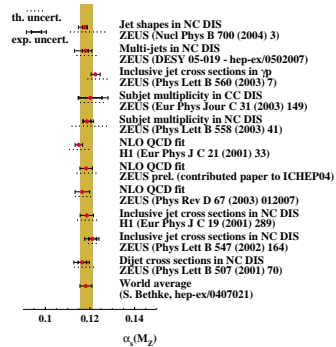, width=0.40\textwidth}
\caption{A summary of the $\as$ from measurements at HERA. \label{alphas_hera}}
\end{wrapfigure}

Figure~\ref{zeus_multijets_subjets} shows the measured dijet and trijet cross sections as a function of $Q^2$ (left)
and their measured ratio $R_{3/2}$ (middle). The NLO predictions of NLOJET ($\mu_R=\mu_F=(\overline{E}_T^2+Q^2)/4$,
CTEQ6) corrected for hadronization effects are compared to the data. NLOJET provides a good
description of both the shape and the magnitude of the measured cross section. The correlated systematic and
the renormalization scale uncertainties largely cancel in the ratio of the cross sections.
This cancellation allows the extraction of $\asz$ with a good precision down to $Q^2$
of $10$~GeV$^2$, using a method similar to that of a previous ZEUS publication~\cite{pl:b507:70}.
The value of $\as$ was measured to be $\asmz{0.1179}{0.0013}{0.0046}{0.0028}{0.0046}{0.0064}$.

Jet substructure for jets produced in DIS has been measured by the ZEUS Collaboration~\cite{zeus_jetsubstructure}
for $Q^2>125$~GeV$^2$. Jets were reconstructed in the laboratory frame and were selected with
$\etjet>17$~GeV and $-1<\etajet<2.5$. The mean integrated jet shape is defined as the averaged
fraction of the jet transverse energy inside the cone of radius $r$:
$\langle\psi(r)\rangle = \frac{1}{N_{\rm jets}} \sum_{\rm jets} \frac{E_T(r)}{\etjet}$,
where $\etjet$ is the total transverse energy of the jet, $E_T(r)$ the part of it
inside the cone of radius $r$ and $N_{\rm jets}$ is the total number of jets in the sample.

Figure~\ref{zeus_multijets_subjets} (right) shows the measured
$\langle\psi(r=0.5)\rangle$ as a function of $\etajet$ and $\etjet$.
There is no significant variation of $\langle\psi(r=0.5)\rangle$ with
$\etajet$ in DIS, whereas $\langle\psi(r=0.5)\rangle$ increases as
$\etjet$ increases.
The sensitivity of the measurements to the value of
$\asz$ is illustrated in Fig.~\ref{zeus_multijets_subjets} (lower part of  plot) by comparing the
measured $\langle\psi(r=0.5)\rangle$ to NLO QCD calculations using
three different values of $\asz$. The NLO QCD calculations provide
a good description of the measured $\langle\psi(r=0.5)\rangle$ and
thus this observable was used to determine $\asz$.
The value of $\asz$ as determined by fitting the NLO QCD calculations
to the measured mean integrated jet shape $\langle\psi(r=0.5)\rangle$
for $\etjet>21$~GeV is 
$\asmz{0.1176}{0.0009}{0.0026}{0.0009}{0.0072}{0.0091}$.

\section{Conclusions}

The HERA measurements show that DGLAP NLO calculations
at low $x$ and forward jet pseudorapidities fail to describe data but the large
theoretical uncertainties prevent a decisive conclusion on parton dynamics at low $x$.
A summary of the $\as$ measurements at HERA is shown in Fig.~\ref{alphas_hera}.
The $\as$ measurements at HERA are in agreement with the world average
and have very competitive errors.
For more accurate measurements of $\as$ improved theoretical calculations would be needed.

\end{document}